\documentclass[12pt]{article}

\usepackage{amsmath}
\usepackage{epsf}

\allowdisplaybreaks

\begin{document}

\baselineskip=18.6pt plus 0.2pt minus 0.1pt

\makeatletter
\@addtoreset{equation}{section}
\renewcommand{\theequation}{\thesection.\arabic{equation}}
\renewcommand{\thefootnote}{\fnsymbol{footnote}}
\newcommand{\nn}{\nonumber}
\newcommand{\bra}[1]{\langle #1\vert}
\newcommand{\ket}[1]{\vert #1\rangle}
\newcommand{\braket}[2]{\langle #1\vert #2\rangle}
\newcommand{\bbbk}[4]{{}_1\langle \, #1 \,|\,{}_2\langle \, #2 \,|\,
                      {}_3\langle \, #3 \,|\,  #4\, \rangle_{123}}
\newcommand{\delB}{\delta_{\text B}}
\newcommand{\QB}{Q_{\text B}}
\newcommand{\tilQ}{\widetilde Q_{\text B}}
\newcommand{\sket}[2][3]{|\, #2\, \rangle_{#1}}
\newcommand{\tket}[1]{|\, #1\, \rangle_{123}}
\newcommand{\bbok}[4]{{}_1\langle \, #1 \,|\,{}_2\langle \, #2 \,|\,
                      #3 \,|\, #4\, \rangle_{123}}
\newcommand{\N}[2]{N^{#1}_{#2}}
\newcommand{\X}[2]{X^{#1}_{#2}}
\newcommand{\inv}[1]{\frac{1}{#1}}
\newcommand{\arn}[2]{\alpha^{(#1)}_{#2}}
\newcommand{\brn}[2]{b^{(#1)}_{#2}}
\newcommand{\crn}[2]{c^{(#1)}_{#2}}
\newcommand{\p}{\partial}
\newcommand{\ap}{\alpha'}
\newcommand{\kp}{\lambda}
\newcommand{\vac}{\ket{0}}
\newcommand{\intund}[1]{\int\nolimits_{#1}}
\newcommand{\Ngh}{N_{\rm gh}}
\newcommand{\Nzz}{N_{00}}
\newcommand{\lnzz}{\ln\!\left(\frac{3^3}{2^4}\right)}
\newcommand{\lnzzsq}{\left[\ln\!\left(\frac{3^3}{2^4}\right)\right]^2}
\newcommand{\delBab}{\delta_{\rm B}^{ab}}
\newcommand{\varphic}{\varphi_{\rm c}}
\newcommand{\bm}[1]{\boldsymbol{#1}}
\newcommand{\y}{y}
\newcommand{\z}{z}
\newcommand{\diag}{\mathop{\rm diag}}
\newcommand{\tv}{non-perturbative vacuum}


\begin{titlepage}
\title{
\hfill\parbox{4cm}
{\normalsize KUNS-1705\\{\tt hep-th/0101162}}\\
\vspace{1cm}
{\Large\bf
Test of the Absence of Kinetic Terms around
the Tachyon Vacuum
in Cubic String Field Theory}
}
\author{
Hiroyuki {\sc Hata}\thanks{{\tt hata@gauge.scphys.kyoto-u.ac.jp}}
{}\hspace*{5pt} and \hspace*{5pt}
Shunsuke {\sc Teraguchi}\thanks{
{\tt teraguch@gauge.scphys.kyoto-u.ac.jp}}
\\[7pt]
{\it Department of Physics, Kyoto University, Kyoto 606-8502, Japan}
}
\date{\normalsize January, 2001}
\maketitle
\thispagestyle{empty}
\begin{abstract}
\normalsize\noindent
It has been conjectured that the bosonic open string theory around
the non-perturbative tachyon vacuum has no open string dynamics
at all.
We explore, in the cubic open string field theory with level
truncation approximation, the possibility that this conjecture is
realized by the absence of kinetic terms of the string field
fluctuations.
We study the kinetic terms with two and four derivatives for the
lower level scalar modes as well as their BRST transformation
properties.
The behavior of the coefficients of the kinetic terms
in the neighborhood of the non-perturbative vacuum supports our
expectation that the BRST invariant scalar component lacks its
kinetic term.

\end{abstract}

\end{titlepage}

\section{Introduction}

The bosonic D-brane has an unstable tachyonic mode on the perturbative
vacuum. It has been conjectured
\cite{Sen:1999mg,Sen:1998sm,Sen:1999mh}
that there is a local minimum of the tachyon potential (\tv),
where the tension of the D-brane and the negative contribution from
the tachyon potential exactly cancel each other. On this vacuum,
there should be no D-branes and hence no open string excitations at
all, leaving only the closed string sector.
It has been also conjectured that the solutions on this
non-perturbative vacuum represent various lower dimensional D-branes
(descent relations).

For studying these conjectures, string field theories have proved to be
very useful, and there have appeared many interesting works in this
direction
\cite{SZ}--\cite{MN}.
In particular, Witten's cubic open string field theory (CSFT)
\cite{Witten:1986cc} combined with the level truncation approximation
\cite{Kostelecky:1988ta} has been successful in verifying the
conjecture on the potential height to a very high degree of accuracy
\cite{SZ,MT}.
Though most of these conjectures, the potential height problem and
the descent relation, have later been proved \cite{GS,KMM} in another
formulation of open string field theory called boundary string field
theory (BSFT) \cite{WittenBSFT1,WittenBSFT2},
there still remains an important problem to be understood: how the
open string excitations disappear around the non-perturbative vacuum.
Note that for analyzing this problem we have to treat all the modes in
open string theory other than the tachyon mode.
Therefore, CSFT whose action is given in a closed form seems to be
rather suited for this problem than BSFT which has concise expression
only for the tachyon field.

One of possible ways to prove the absence of physical open string
excitations is to show that they all have infinite masses \cite{T2}.
Another and roughly an equivalent way is to show that the physical
components of the open string field fluctuations lack their kinetic
terms in the SFT action and hence become non-dynamical
\cite{Sen:1999md} (kinetic term
means the quadratic term of the fluctuations containing the
derivatives with respect to the center-of-mass string coordinate).
In relation to this, an interesting deformed version of CSFT has
recently been proposed in \cite{RSZ} as a possible SFT model around
the non-perturbative vacuum. The BRST operator in this model is simply
equal to the ghost zero-mode, $\QB=c_0$, and hence the quadratic term
$\Phi\QB\Phi$ contains no kinetic term at all.
What we have to do for proving the absence of open string excitations
on the non-perturbative vacuum is to derive this kind of SFT from the
ordinary CSFT.

The purpose of this paper is to explore, using the level truncation
approximation, the possibility that CSFT action expanded around the
non-perturbative vacuum contains no kinetic terms for physical
fluctuations. Let the action of CSFT be given by
$S=(1/2)\,\Phi\,\QB\Phi +(1/3)\,\Phi^3$, and let us expand
the original string field $\Phi$ as $\Phi=\Phi_0 + \Psi$ with $\Phi_0$
being the classical solution representing the non-perturbative vacuum.
Then the term in $S$ quadratic in the fluctuation $\Psi$ is
$(1/2)\Psi(\QB + 2\,\Phi_0\star)\Psi$.
Since the three-string interaction vertex defining the star-product
$\star$ contains any powers of the derivative $\p/\p x^\mu$ with
respect to the center-of-mass $x^\mu$ of the string coordinate
$X^\mu(\sigma)$,
the kinetic term for the fluctuation $\Psi$ consists of terms with any
number of derivatives. Our expectation is that these kinetic terms all
vanish for the physical components of $\Psi$, where ``physical'' means
being BRST invariant.
We are going to study this possibility in the level (2,6) truncation by
taking as the fluctuations only the scalar modes at levels zero and
two. As the kinetic terms we take the quadratic and the quartic ones
in the derivatives $\p/\p x^\mu$.

As the simplest and encouraging example, let us quote our result in
the level $(0,0)$ approximation, namely, let us take only the level
zero scalar field $t$ (see sec.\ 3.1). The kinetic term for $t$
quadratic in the derivative is
\begin{equation}
{\cal L}_{\rm kin}=
-\frac12\left[1 - \frac{81\sqrt{3}}{32}\lnzz t_0\right]
\left(\p_\mu t\right)^2 ,
\label{eq:Skin}
\end{equation}
where $t_0$ is the classical value of $t$ at the non-perturbative
vacuum. This kinetic term vanishes at
$t_0=32/(81\sqrt{3}\ln(3^3/2^4))\simeq 0 .436$, which is close
to the value of $t_0$ obtained as the stationary point of the level
(0,0) potential, $t_0\simeq 0.456$.
Even this simplest example supports our expectation that the kinetic
term for the physical fluctuations totally vanish.

In our analysis in the level (2,6) approximation, we consider the
kinetic terms quadratic and quartic in
the derivatives for the three scalar fluctuation modes at levels zero
and two. Then the coefficients of the kinetic terms form $3\times 3$
matrices in the space of fluctuations. Since there is only one
physical component among the three scalar fluctuations, the two
matrices, one for the kinetic term with two derivatives and
the other with four, must have a common zero-mode (namely, the
eigenvector corresponding to the eigenvalue zero) in order for the
physical fluctuation to disappear from the kinetic terms.
We examine numerically whether this condition is satisfied in the
neighborhood of the non-perturbative vacuum, and obtain results which
support our expectation.

The organization of the rest of this paper is as follows.
In sec.\ 2, after making a setup for the level truncation calculation
in CSFT, we present the derivative expansion of the kinetic term of
the action and the BRST transformation at a generic translation
invariant background.
Then, in sec.\ 3 we carry out the numerical analysis of the kinetic
terms and the BRST transformations of the fluctuations to examine
whether the kinetic terms for the physical fluctuation can disappear
in the neighborhood of the non-perturbative vacuum.
Finally in sec.\ 4, we discuss future problems.

\section{Derivative expansion of the SFT action}

In this section, as a preparation for the analysis in sec.\ 3, we
fix our conventions for the CSFT action, and then present
the derivative expansion of the kinetic terms and the BRST
transformation for the scalar fluctuation modes in the level (2,6)
approximation.

\subsection{SFT action and the BRST transformation}

We shall first recapitulate the gauge-fixed action and the BRST
transformation of CSFT in the Siegel gauge.
They are both obtained from the gauge invariant action:
\begin{equation}
S_{\rm inv} =-\frac{1}{g_o^2}\left(
\frac12\intund{b_0,x}\bra{\Phi}\QB\ket{\Phi}+ \frac13
\intund{b_0,x}^{(3)}\intund{b_0,x}^{(2)}\intund{b_0,x}^{(1)}
\bbbk{\Phi}{\Phi}{\Phi}{V}\right) ,
\label{eq:Sinv}
\end{equation}
where $g_o$ is the open string coupling constant, and
$\intund{b_0,x}^{(r)}\equiv\int db_0^{(r)}\int d^{26}x_r$ denotes the
integrations over the anti-ghost zero-mode $b_0$ and the
center-of-mass coordinate $x^\mu$ of the $r$-th string.
The string field $\ket{\Phi}$ in (\ref{eq:Sinv}) is a state
in the Fock space of the first quantized string and carry ghost number
$-1$. It is expanded in terms of the anti-ghost zero mode $b_0$ as
\begin{equation}
\ket{\Phi(x,b_0)} = b_0 \ket{\phi(x)} + \ket{\psi(x)} ,
\label{eq:Phi=b0phi+psi}
\end{equation}
where $\ket{\phi}$ and $\ket{\psi}$ carry ghost number $0$ and $-1$,
respectively.
The ghost zero-mode structure of the BRST operator $\QB$ is
\begin{align}
\QB & = c_0 L + b_0 M + \tilQ,
\end{align}
where we have
\begin{align}
L \;\, & = \frac12\,\alpha_0^2 +\sum_{n=1}^{\infty}\left(
\alpha_{-n}\cdot\alpha_{n} + n c_{-n}b_{n} + n b_{-n}c_{n} \right) -1,\\
M \; & = -2 \sum_{n=1}^{\infty} n c_{-n}c_{n},\\
\tilQ & = \sum_{n\neq 0}c_{-n} \sum_{m=-\infty}^{\infty}\frac12\,
            \alpha_{n-m}\cdot\alpha_{m}
            + \sum_{\substack{n\neq 0, m\neq 0\\  n+m\neq 0 }}
\frac{m-n}{2}\, c_m c_n b_{-m-n} ,
\end{align}
with $(\alpha_0)_\mu =\sqrt{2}\,p_\mu=\sqrt{2}\,(-i\p/\p x^\mu)$.
The three-string interaction vertex $\tket{V}$ is given in the momentum
representation for $x^\mu$ as
\begin{align}
\tket{V} & = \kp\,\exp\left[\sum_{r,s=1}^{3}
\left( \inv{2}\sum_{n,m=0}^{\infty}
N^{rs}_{nm}\arn{r}{-n}\cdot\arn{s}{-m}
- \sum_{n=1}^{\infty}\sum_{m=0}^\infty
X^{rs}_{nm}\crn{r}{-n}\brn{s}{-m} \right)
\right]\tket{0} \nn\\
&\hspace*{5cm}\times (2\pi)^{26}\delta^{26}\left(p_1+p_2+p_3\right) ,
\end{align}
where $\vac$ is the Fock vacuum satisfying $(\alpha_n,b_n,c_n)\vac=0$
($n\ge 1$), $\kp$ is a constant $\kp=3^{9/2}/2^6$ \cite{SZ}, and the
Neumann coefficients, $N_{nm}^{rs}$ and $X_{nm}^{rs}$, are defined in
\cite{T,HS}. In particular, for $N_{00}^{rs}$ we have
$\sum_{r,s=1}^3N_{00}^{rs}\alpha_0^{(r)}\cdot\alpha_0^{(s)}
=N_{00}\sum_{r=1}^3\bigl(\alpha_0^{(r)}\bigr)^2$
with $N_{00}= - (1/2)\ln(3^3/2^4)$.

Then the gauge-fixed action $S$ in the Siegel gauge,
$b_0\ket{\Phi}=0$, is obtained from $S_{\rm inv}$ (\ref{eq:Sinv})
simply by putting $\psi=0$:
\begin{equation}
S=-\frac{1}{g_o^2}\left(
\frac12\intund{x}\bra{\phi}L\ket{\phi}
+ \frac13\intund{x_1}\intund{x_2}\intund{x_3}
\bbbk{\phi}{\phi}{\phi}{v}
\right) ,
\label{eq:S}
\end{equation}
where the three-string vertex for the gauge-fixed action is
$\tket{v}\equiv\tket{V}\vert_{b_0^{(1)}=b_0^{(2)}=b_0^{(3)}=0}$.
This action has an invariance under the BRST transformation
$\delB\phi\propto(\delta S_{\rm inv}/\delta\psi)_{\psi=0}$:
\begin{equation}
i\,\delB \sket{\phi} = \tilQ^{(3)} \sket{\phi} -
\intund{x_1}\intund{x_2}
\sum_{r=1}^{3}\sum_{n=1}^{\infty}\X{r3}{n0}\;
\bbok{\phi}{\phi}{c_{-n}^{(r)}}{v} \ .
\label{eq:delB}
\end{equation}
Physical quantities in the present SFT are required to be invariant
under the BRST transformation.

\subsection{Derivative expansion of the action}

As mentioned in sec.\ 1, we are interested in the kinetic term of the
string field fluctuation around the \tv.
To analyze it in the simplest framework, we adopt the level (2,6)
truncation and take the scalar fields at levels zero and two,
both as the vacuum coordinates and the fluctuations.
Namely, we restrict the $\Ngh=0$ component field sector of
$\ket{\phi}$ to the following form:
\begin{equation}
\ket{\phi(x)}_{\Ngh=0}=  t(x)\,\vac +  u(x)\,c_{-1}b_{-1}\vac
+v(x)\,\frac{1}{\sqrt{52}}\left(\alpha_{-1}\cdot\alpha_{-1}
\right)\vac ,
\end{equation}
Then, reexpressing the three component field
$\left(\varphi^i\right)=\left(t,u,v\right)$ as
$\varphi^i\to\varphi^i_0 + \varphi^i$ with $\varphi^i_0$ being a
(translation invariant) vacuum value and $\varphi^i$ the fluctuation
around $\varphi^i_0$, we carry out the derivative expansion of the
kinetic term of the fluctuation
$\varphi^i$ obtained from the action $S$ (\ref{eq:S}):
\begin{equation}
S_{\rm kin}=-\frac{1}{2\,g_o^2}\int\!d^{26}x\left(
G_{ij}(\varphi_0)\,\p_\mu\varphi^i\,\p_\mu\varphi^j
+ H_{ij}(\varphi_0)\,\p^2\varphi^i\,\p^2\varphi^j + \ldots
\right) .
\label{eq:expandS}
\end{equation}

Here we are making the simplification of considering only the scalar
fields $(t,u,v)$.
In particular, we took only the trace part of the tensor fields
$v_{\mu\nu}$ which appear in the expansion of $\ket{\phi}$ in the form
$v_{\mu\nu}(x)\,\alpha_{-1}^\mu\alpha_{-1}^\nu\vac$. We have also
discarded the vector field at level two which appear as the
coefficient of $\alpha_{-2}^\mu\vac$.
Complete treatment in the (2,6) truncation should take into account the
mixing among all the scalar, vector and tensor fields at levels zero,
one and two. However, such analysis would be so complicated since the
size of the matrices $G_{ij}$ and $H_{ij}$ become much larger, and
there are many physical fields that should be confined.
Therefore, we shall carry out the following analysis in the smallest
fluctuation space $(t,u,v)$ with a hope that the mixing with other
components is small.

Now each component of the $3\times 3$ matrix $G_{ij}(\varphi_0)$ is
a linear function of $\varphi_0$ and is given explicitly by
(hereafter we omit the subscript 0 on $\varphi_0$ unless confusion
occurs)
\begin{align}
&G_{tt}= 1+\frac32\sqrt{\frac{3}{13}}\,v
+\frac{\sqrt{3}}{32}\lnzz\left(-81\,t-33\,u+15\sqrt{13}\,v\right) ,
\label{eq:Gtt}
\\
&G_{uu}= -1+\frac{19}{54\sqrt{39}}\,v
+\frac{\sqrt{3}}{32}\lnzz\left(
-\frac{19}{3}\,t -u + \frac{95\sqrt{13}}{81}\,v\right) ,
\\
&G_{vv}= 1-\frac{49}{156\sqrt{3}}\,t -\frac{539}{4212\sqrt{3}}\,u
+\frac{2779}{468\sqrt{39}}\,v
+\frac{\sqrt{3}}{32}\lnzz\left(
-\frac{581}{9}\,t -\frac{6391}{243}\,u +\frac{20951}{81\sqrt{13}}\,v
\right) ,
\\
&G_{tu}= \frac{11}{6\sqrt{39}}\,v +\frac{\sqrt{3}}{32}\lnzz\left(
-33\,t -\frac{19}{3}\,u +\frac{55\sqrt{13}}{9}\,v\right) ,
\\
&G_{uv}= \frac{11}{24\sqrt{39}}\,t +\frac{19}{216\sqrt{39}}\,u
-\frac{4279}{8424\sqrt{3}}\,v + \frac{\sqrt{3}}{32}\lnzz\left(
\frac{55\sqrt{13}}{9}\,t +\frac{95\sqrt{13}}{81}\,u
-\frac{6391}{243}\,v\right) ,
\\
&G_{vt}=\frac{3}{8}\sqrt{\frac{3}{13}}\,t +\frac{11}{24\sqrt{39}}\,u
-\frac{389}{312\sqrt{3}}\,v + \frac{\sqrt{3}}{32}\lnzz\left(
15\sqrt{13}\,t +\frac{55\sqrt{13}}{9}\,u -\frac{581}{9}\,v\right) .
\label{eq:Gvt}
\end{align}
Note that in the present convention we have
$G_{ij}(\varphi^i=0)=\diag\left(1,-1,1\right)$ at the perturbative
vacuum (the minus sign for the $u$ field is due to the
negative norm property of the ghosty state $c_{-1}b_{-1}\vac$).

The matrix $H_{ij}(\varphi)$ for the four-derivative
kinetic term have contributions only from the interaction term of
(\ref{eq:S}). It is given by
\begin{align}
H_{tt}=&-\frac32\sqrt{\frac{3}{13}}\lnzz v
+ \frac{3\sqrt{3}}{64}\lnzzsq\left(
27\,t + 11\,u - 5\sqrt{13}\,v\right) ,
\\
H_{uu}=&-\frac{19}{54\sqrt{39}}\lnzz v
+\frac{\sqrt{3}}{64}\lnzzsq\left(
\frac{19}{3}\,t + u - \frac{95\sqrt{13}}{81}\,v\right) ,
\\
H_{vv}=&\frac{1}{\sqrt{3}}\left(\frac{1}{78}\,t +\frac{11}{2106}\,u
-\frac{5}{18\sqrt{13}}\,v\right)
+\frac{1}{156\sqrt{3}}\lnzz\left(
49\,t + \frac{539}{27}\,u -\frac{2779}{3\sqrt{13}}\,v\right)
\nn\\
&+\frac{1}{192\sqrt{3}}\lnzzsq\left(
581\,t +\frac{6391}{27}\,u -\frac{20951}{9\sqrt{13}}\,v\right) ,
\\
H_{tu}=& -\frac{11}{6\sqrt{39}}\lnzz v
+ \frac{\sqrt{3}}{64}\lnzzsq\left(
33\,t +\frac{19}{3}\,u -\frac{55\sqrt{13}}{9}\,v\right) ,
\\
H_{uv}=& \frac{22}{1053\sqrt{3}}\,v
+\frac{1}{24\sqrt{3}}\lnzz\left(-\frac{11}{\sqrt{13}}\,t
-\frac{19}{9\sqrt{13}}\,u +\frac{4279}{351}\,v\right)
\nn\\
&+\frac{1}{192\sqrt{3}}\lnzzsq\left(
-55\sqrt{13}\,t - \frac{95\sqrt{13}}{9}\,u +\frac{6391}{27}\,v
\right) ,
\\
H_{vt}=& \frac{2}{39\sqrt{3}}\,v
+\frac{\sqrt{3}}{8}\lnzz\left(-\frac{3}{\sqrt{13}}\,t
-\frac{11}{9\sqrt{13}}\,u +\frac{389}{117}\,v\right)
\nn\\
&+\frac{\sqrt{3}}{64}\lnzzsq\left(-15\sqrt{13}\,t
-\frac{55\sqrt{13}}{9}\,u +\frac{581}{9}\,v\right) .
\end{align}

For completeness we present the potential $V(\varphi)$ in the (2,6)
truncation:
\begin{align}
V(\varphi) =&
- \frac{1}{2}\,t^2 + \frac{27\sqrt{3}}{64}\,t^3
- \frac{1}{2}\,u^2 + \frac{1}{2}\,v^2
+ \frac{33\sqrt{3}}{64}\,t^2\,u
- \frac{15\sqrt{39}}{64}\,t^2\,v
\nn\\
&+ \frac{19}{64\sqrt{3}}\,t\,u^2
- \frac{55}{96}\sqrt{\frac{13}{3}}\,t\,u\,v
+ \frac{581}{192\sqrt{3}}\,t\,v^2
\nn\\
&+ \frac{1}{64\sqrt{3}}\,u^3
- \frac{95}{1728}\sqrt{\frac{13}{3}}\,u^2\,v
+ \frac{6391}{5184\sqrt{3}}\,u\,v^2
- \frac{20951}{5184\sqrt{39}}\,v^3 \ .
\label{eq:26V}
\end{align}

\subsection{Derivative expansion of the BRST transformation}

In the analysis of sec.\ 3, we also need the derivative expansion of
the BRST transformation of the fields $(t,u,v)$ for identifying their
physical (BRST invariant) combination at the \tv.
Since the ghost field
in $\ket{\phi}$ at level one does not contribute to $\delB(t,u,v)$,
we have only to consider the ghost fields at level two, $C(x)$ and
$C_\mu(x)$, which appear in $\ket{\phi}$ as
\begin{equation}
\ket{\phi(x)}_{\mbox{\scriptsize level-2 ghost sector}}=
i\,C(x)\,b_{-2}\vac + C_\mu(x)\,\alpha_{-1}^\mu b_{-1}\vac .
\end{equation}
Then, $\delB(t,u,v)$ can be calculated using the formula
(\ref{eq:delB}). Similarly to the kinetic term (\ref{eq:expandS}), we
carry out the derivative expansion for the contribution from the last
term of (\ref{eq:delB}) and express the result as
\begin{equation}
\delB\varphi^i = a^i(\varphi_0)\,C + b^i(\varphi_0)\,\p_\mu C_\mu
+ \y^i(\varphi_0)\,\p^2 C + z^i(\varphi_0)\,\p^2\p_\mu C_\mu
+ \ldots ,
\label{eq:expanddelB}
\end{equation}
where we have omitted those terms containing the fluctuation
$\varphi^i$ on the RHS.
The explicit expression of the coefficients
$\left(a^i(\varphi),b^i(\varphi),\y^i(\varphi),z^i(\varphi)\right)$
are as follows:
\begin{align}
a^t=&\frac{3\sqrt{3}}{4}\,t - \frac{29}{12\sqrt{3}}\,u
- \frac{5}{12}\sqrt{\frac{13}{3}}\,v ,
\\
a^u=& -3 - \frac{11}{12\sqrt{3}}\,t +\frac{703}{324\sqrt{3}}\,u
+ \frac{55}{324}\sqrt{\frac{13}{3}}\,v ,
\\
a^v=& -\sqrt{13} -\frac{5}{12}\sqrt{\frac{13}{3}}\,t
+\frac{145}{324}\sqrt{\frac{13}{3}}\,u +\frac{581}{324\sqrt{3}}\,v ,
\\
b^t=& \frac{3}{2}\sqrt{\frac{3}{2}}\,t
+\frac{19}{6\sqrt{6}}\,u -\frac{97}{6\sqrt{78}}\,v ,
\\
b^u=& \sqrt{2} - \frac{11}{6\sqrt{6}}\,t
- \frac{1}{2\sqrt{6}}\,u + \frac{1067}{162\sqrt{78}}\,v ,
\\
b^v=& \sqrt{\frac{2}{13}} -\frac{49}{6\sqrt{78}}\,t
-\frac{931}{162\sqrt{78}}\,u +\frac{2779}{702\sqrt{6}}\,v ,
\\
\y^t=& -\frac{4}{3\sqrt{39}}\,v +\frac{1}{12\sqrt{3}}\lnzz\left(
27\,t -29\,u -5\sqrt{13}\,v\right) ,
\\
\y^u=& \frac{44}{81\sqrt{39}}\,v
+ \frac{1}{12\sqrt{3}}\lnzz\left(
-11\,t +\frac{703}{27}\,u +\frac{55\sqrt{13}}{27}\,v\right) ,
\\
\y^v=& -\frac{1}{3\sqrt{39}}\,t +\frac{29}{81\sqrt{39}}\,u
+ \frac{389}{1053\sqrt{3}}\,v
+\frac{1}{12\sqrt{3}}\lnzz\left(
-5\sqrt{13}\,t +\frac{145\sqrt{13}}{27}\,u +\frac{581}{27}\,v\right) ,
\\
z^t=& -\frac{4}{3}\sqrt{\frac{2}{39}}\,v
+ \frac{1}{6\sqrt{6}}\lnzz\left(
27\,t + 19\,u -\frac{97}{\sqrt{13}}\,v\right) ,
\\
z^u=& \frac{44}{81}\sqrt{\frac{2}{39}}\,v
+ \frac{1}{6\sqrt{6}}\lnzz\left(
-11\,t - 3\,u +\frac{1067}{27\sqrt{13}}\,v\right) ,
\\
z^v=& -\frac{1}{3}\sqrt{\frac{2}{39}}\,t
-\frac{19}{81}\sqrt{\frac{2}{39}}\,u
+\frac{119}{351}\sqrt{\frac{2}{3}}\,v
+\frac{1}{6\sqrt{6}}\lnzz\left(-\frac{49}{\sqrt{13}}\,t
-\frac{931}{27\sqrt{13}}\,u +\frac{2779}{117}\,v\right) .
\end{align}
Note that the constant terms in $a^u$, $a^v$, $b^u$ and $b^v$ came
from $\tilQ\ket{\phi}$ in (\ref{eq:delB}), and that the tachyon
field $t$ is BRST invariant at the perturbative vacuum.

\section{Testing the absence of the kinetic terms}

In this section, we shall study numerically whether the conditions for
the absence of the kinetic terms of the physical scalar field are
realized near the \tv.
The precise expressions of the conditions will be stated below,
in particular, in sec.\ 3.2.
What we shall test are
i) location of the zero of $\det G$ and that of $\det H$,
ii) coincidence of the zero-modes of $G_{ij}$ and $H_{ij}$,
and iii) existence of the BRST invariant linear combination of
$(t,u,v)$ including the higher derivative terms in $\delB\varphi^i$.
Unfortunately, these conditions are not satisfied to high precision at
the \tv\ determined as a stationary point of the potential.
Therefore, we shall look at each of them globally in the space
of vacuum coordinate $\varphi_0^i=(t_0,u_0,v_0)$ to test whether
there is a point satisfying the condition near the \tv.

\subsection{Analysis at level (0,0)}
Let us first consider the simplest (0,0) truncation, namely, let us
keep only the level zero field $t$.
In this approximation the field $t$ is physical everywhere,
$\delB^{(0,0)} t=0$, and the potential $V$ (\ref{eq:26V}) and the
metric $G_{tt}$ (\ref{eq:Gtt}) are reduced to
\begin{align}
&V^{(0,0)} = -\frac12\,t^2 + \frac{27\sqrt{3}}{64}\,t^3 ,
\label{eq:V00}
\\
&G_{tt}^{(0,0)}=1 - \frac{81\sqrt{3}}{32}\lnzz t \ .
\label{eq:G00}
\end{align}
The non-trivial stationary point (\tv) of $V^{(0,0)}$
exists at $t=64/(81\sqrt{3})\simeq 0.456$,
while the metric $G_{tt}^{(0,0)}$ vanishes at
$t=32/(81\sqrt{3}\ln(3^3/2^4))\simeq 0 .436$.
Surprisingly these two points are very close, supporting that the
kinetic term of the physical field $t$ is missing.
As for $H_{tt}$ for the four-derivative kinetic term, it vanishes only
at $t=0$. However, the level (0,0) approximation would be too poor for
such higher derivative kinetic terms.

\subsection{Analysis of $\bm{G_{ij}}$}

Encouraged by the above result in the (0,0) approximation, let us
proceed to the analysis of the full $G_{ij}$ of eqs.\ (\ref{eq:Gtt})
-- (\ref{eq:Gvt}) and the first two terms of the BRST transformation
(\ref{eq:expanddelB}):
\begin{equation}
\delBab\varphi^i = a^i(\varphi)\,C + b^i(\varphi)\,\p_\mu C_\mu \ .
\label{eq:expanddelBab}
\end{equation}
Namely, we shall examine the lowest derivative kinetic term and the
lowest derivative part of the BRST transformation.

Before starting the analysis,
let us state our expectation for the disappearance of the
physical open string modes in the present situation.
Since we have two independent ghost fields, $C$ and $\p_\mu C_\mu$, on
the RHS of (\ref{eq:expanddelBab}), there is only one BRST invariant
linear combination $T$ of the three scalar fields $(t,u,v)$ satisfying
$\delBab\,T=0$:
\begin{equation}
T= t + \beta_u(\varphi)\,u + \beta_v(\varphi)\,v ,
\label{eq:T}
\end{equation}
where $\beta_i(\varphi)$ ($i=u,v$) is the solution of
\begin{align}
&a^t + \beta_u\,a^u + \beta_v\,a^v =0 ,
\nn\\
&b^t + \beta_u\,b^u + \beta_v\,b^v =0 .
\label{eq:eqforbeta}
\end{align}
Namely, the level zero scalar $t$, which was BRST invariant at the
perturbative vacuum, needs a mixing of $u$ and $v$ to reconstruct a
BRST invariant field at the \tv.

A possible mechanism that confines this physical scalar $T$ is that
$T$ has no kinetic term. In particular, $T$ should not appear in
$G_{ij}\p_\mu\varphi^i\p_\mu\varphi^j$.
For this to be realized, it is sufficient that the matrix
$G_{ij}(\varphi)$ has a zero-mode $\bm{e}_0$ (i.e., the eigenvector of the
eigenvalue zero) and that it has a non-vanishing inner product with
$\bm{\beta}\equiv (1,\beta_u,\beta_v)$:
\begin{equation}
\bm{\beta}\cdot\bm{e}_0 \ne 0 .
\label{eq:betae0}
\end{equation}
The reason is as follows. Let $\bm{e}_a$ ($a=0,1,2$) be the three
orthonormal eigenvectors of the (real and symmetric) matrix $G_{ij}$
corresponding to the eigenvalue $g_a$:
\begin{equation}
G\,\bm{e}_a=g_a\,\bm{e}_a,\quad
\bm{e}_a\cdot\bm{e}_b=\delta_{a,b},\quad
g_0=0 .
\label{eq:ea}
\end{equation}
Then, the two-derivative kinetic term is diagonalized as
\begin{equation}
G_{ij}\p_\mu\varphi^i\p_\mu\varphi^j=
\sum_{a=1,2}g_a\left(\p_\mu\phi_a\right)^2 ,
\label{eq:kineticpsi}
\end{equation}
where the new field $\phi_a$ is defined by
$\phi_a\equiv\bm{e}_a\cdot\bm{\varphi}$.
In order for the kinetic term (\ref{eq:kineticpsi}) not to contain the
physical $T=\bm{\beta}\cdot\bm{\varphi}$, $T$ should not be expressed
as a linear combination of $\phi_a$ ($a=1,2)$, and this is realized if
the condition (\ref{eq:betae0}) holds.
We would like to emphasize that the missing combination
$\bm{e}_0\cdot\bm{\varphi}$ need not be exactly equal to
$T$.\footnote{
The requirement $\bm{\beta}\propto \bm{e}_0$ is not
invariant under a transformation of the coordinate system
$\varphi^i$ of the field space.
}
A rather weak condition (\ref{eq:betae0}) is sufficient.

Therefore, what we have to test first of all is whether the
determinant of the matrix $G_{ij}(\varphi)$ vanishes near the \tv.
Recall that the coordinate $\varphi_c=(t_c,u_c,v_c)$ of the \tv\
obtained as a stationary point of the level (2,6) potential
(\ref{eq:26V}) is \cite{SZ}
\begin{equation}
t_c\simeq 0.544,\quad
u_c\simeq 0.190,\quad
v_c\simeq 0.202 .
\label{eq:varphi_c}
\end{equation}
First, the value of $\det G$ at the point of (\ref{eq:varphi_c}) is
$\det G(0.544,0.190,0.202)=0.189$, which is relatively close to
zero compared with its value at the perturbative vacuum $\det
G(0,0,0)=-1$, indicating that the point (\ref{eq:varphi_c}) is near
the two-dimensional surface of $\det G=0$, though not exactly on it.

\begin{figure}[htb]
\begin{center}
\leavevmode
\epsfxsize=110mm
\put(317,148){{\Large $\bm{t}$}}
\put(20,258){{\Large $\bm{\det G}$}}
\put(290,253){($0,0$)}
\put(310,238){($0.1,0.1$)}
\put(310,225){($0.2,0.2$)}
\put(290,200){$(0.3,0.3)$}
\epsfbox{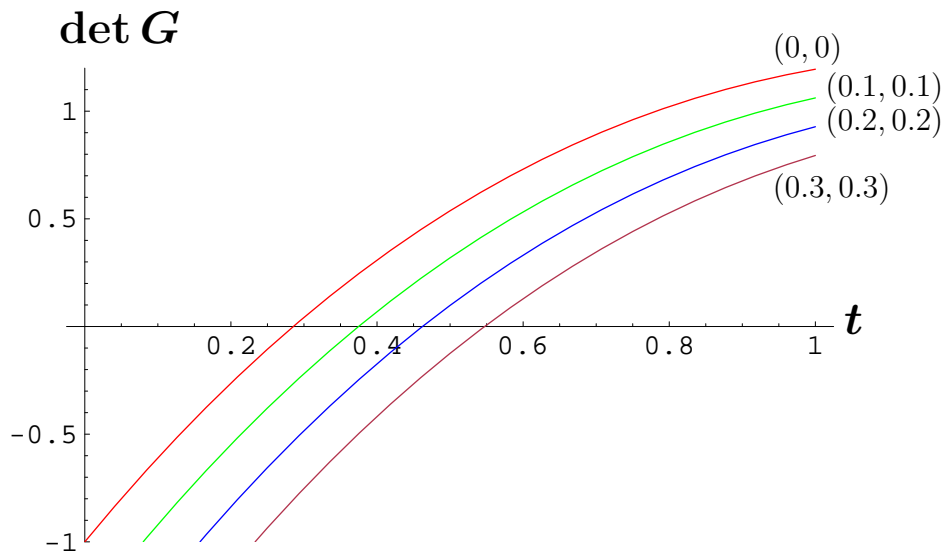}
\vspace{-2cm}
\caption{
$\det G$ as a function of $t$ for
$(u,v)=(0,0),(0.1,0.1),(0.2,0.2),(0.3,0.3)$.
}
\label{fig:plotdetg}
\end{center}
\end{figure}

\begin{figure}[htb]
\begin{center}
\leavevmode
\epsfxsize=100mm
\put(290,120){{\Large $\bm{u}$}}
\put(21,230){{\Large $\bm{v}$}}
\put(285,220){$t=0.60$}
\put(285,190){$t=0.55$}
\put(285,163){$t=0.50$}
\put(285,140){$t=0.45$}
\put(285,105){$t=0.40$}
\epsfbox{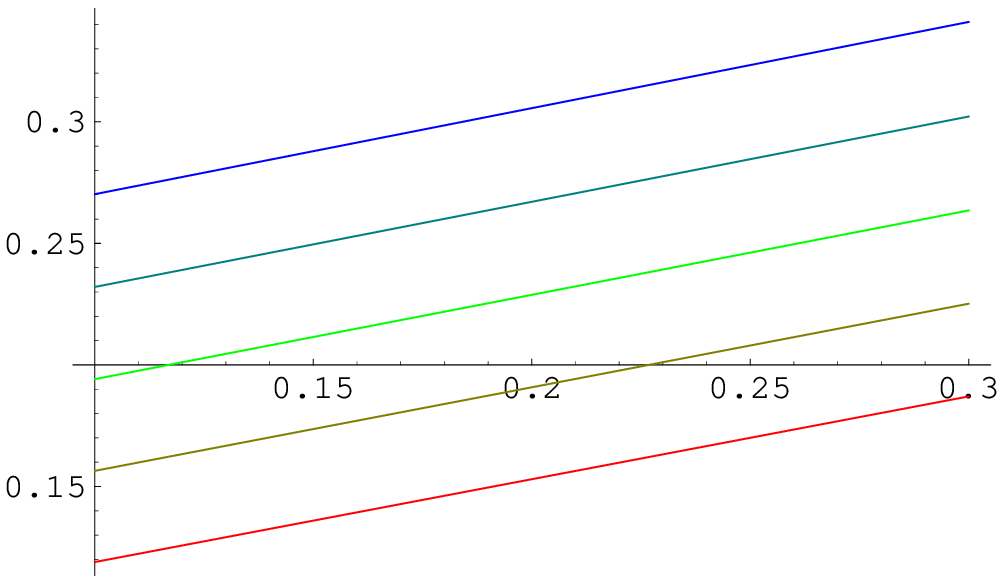}
\vspace{-2cm}
\caption{
Curves of $\det G=0$ in the $(u,v)$ plane for $t=0.4,0.45,0.5,0.55,0.6$.
}
\label{fig:detgzerocurve}
\end{center}
\end{figure}

Figs.\ \ref{fig:plotdetg} and \ref{fig:detgzerocurve} would help
understanding how the \tv\ is close to the surface $\det G=0$.
In fig.\ \ref{fig:plotdetg} we plot $\det G$ as
a function of $t$ for various fixed values of $(u,v)$.
In particular, the zero of $\det G$ for $(u,v)=(u_c,v_c)$ of
(\ref{eq:varphi_c}) is at $t=0.469$, while the zero for $(u,v)=(0,0)$
is at $t=0.285$.\footnote{
For $(u,v)=(u_c,v_c)$, $\det G$ has other two zeros at $t=2.82$ and
$3.81$. When $(u,v)=(0,0)$, the other two zeros are complex.
}
On the other hand,
fig.\ \ref{fig:detgzerocurve} shows curves of $\det G=0$ in the
$(u,v)$ plane for several fixed values of $t$.
As a concrete example, assuming the conjectured equality $u=v$ at the
\tv\ \cite{SZ,HS}, the value of $u$ satisfying
$\det G(t_c=0.544,u,u)=0$ is $u=0.296$.

As a representative point on the $\det G=0$ surface which is
reasonably close to the \tv\ of (\ref{eq:varphi_c}), there is a
stationary point $(t_r,u_r,v_r)$ of the level (2,6) potential
(\ref{eq:26V}) restricted on the surface $\det G=0$:
\begin{equation}
(t_r,u_r,v_r)\simeq (0.500,0.181,0.222) .
\label{eq:varphi_r}
\end{equation}
The value of the potential (\ref{eq:26V}) (normalized so that the true
value is equal to $-1$) at this point is
$2\pi^2 V(t_r,u_r,v_r)=-0.904$, while that at the point
(\ref{eq:varphi_c}) is $2\pi^2 V(t_c,u_c,v_c)=-0.959$.

The above analysis tells that the \tv\ is indeed close to
the surface $\det G=0$. In particular, for $t\sim 0.5$ the condition
$\det G=0$ combined with the assumption $u\sim v$
predicts the expected value $\sim 0.2$ for $u$ and $v$.
We have also examined the condition (\ref{eq:betae0}) to find that
the inner product $\bm{\beta}\cdot\bm{e}_0$ on the surface $\det G=0$
has no zeros at least in the region $0<u,v<1$.

\subsection{Analysis of $\bm{H_{ij}}$}

Next let us consider the four-derivative kinetic term in
(\ref{eq:expandS}).
What we expect for $H_{ij}(\varphi)$ in order for the physical scalar
$T$ (\ref{eq:T}) to drop out from the four-derivative kinetic term is
as follows:
\begin{itemize}
\item $H_{ij}$ has a zero-mode $\bm{f}_0$ at the \tv.
\item The two non-zero-modes of $H_{ij}$ span the
same two-dimensional subspace as that spanned by the non-zero-modes of
$G_{ij}$ (equivalently, $\bm{f}_0$ is parallel to the zero-mode
$\bm{e}_0$ of $G_{ij}$).

\end{itemize}
If these conditions are satisfied, then
$H_{ij}\p^2\varphi^i\p^2\varphi^j$ is also expressed only in terms of
$\phi_a$ ($a=1,2$) appearing in the two-derivative kinetic term
(\ref{eq:kineticpsi}).

\begin{figure}[htb]
\begin{center}
\leavevmode
\epsfxsize=110mm
\put(317,175){{\Large $\bm{t}$}}
\put(0,258){{\Large $\bm{10^5\det H}$}}
\put(106,55){($0,0$)}
\put(157,75){($u_r,v_r$)}
\put(210,55){($0.15,0.30$)}
\put(275,75){($0.15,0.35$)}
\epsfbox{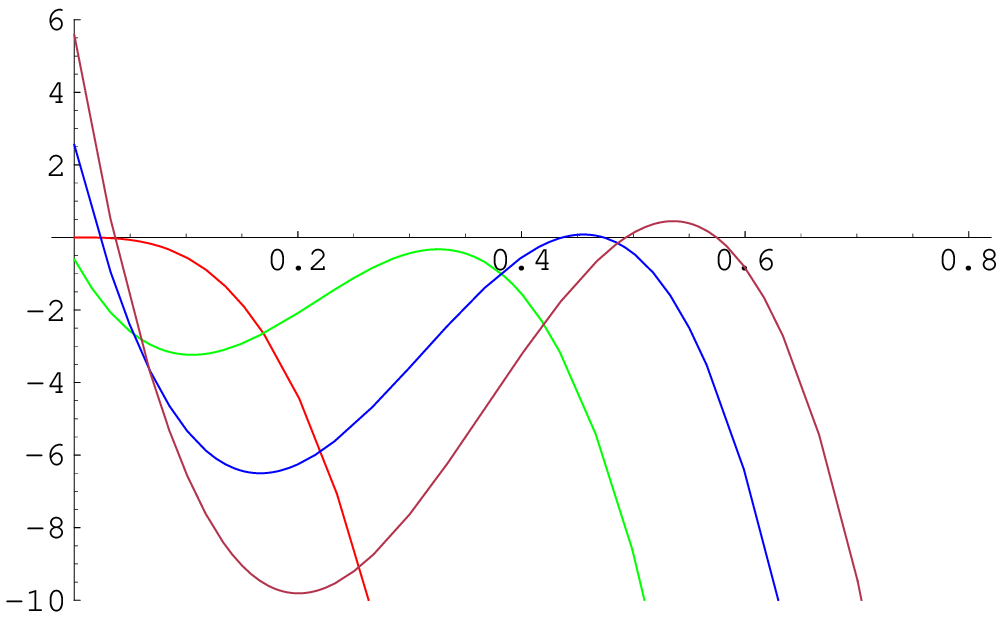}
\vspace{-1.5cm}
\caption{
$\det H$ as a function of $t$ for
$(u,v)=(0,0),(u_r,v_r),(0.15,0.30),(0.15,0.35)$.
}
\label{fig:plotdetH}
\end{center}
\end{figure}

First, let us see whether $\det H$ vanishes near the \tv.
Fig.\ \ref{fig:plotdetH} shows $\det H$ as a function of $t$
for several fixed values of $(u,v)$.
For $(u,v)=(0,0)$, $\det H$ has no zeros in the range of $t$ of our
interest, and for $(u,v)=(u_r,v_r)$ of (\ref{eq:varphi_r}), $\det H$
become close to zero at $t\sim 0.33$ but does not cross zero.
For larger values of $v$ compared with $u$, $(u,v)=(0.15,0.30)$ and
$(0.15,0.35)$, $\det H$ has zeros around $t=0.5$.

\begin{figure}[htb]
\begin{center}
\leavevmode
\epsfxsize=100mm
\put(290,140){{\Large $\bm{v}$}}
\put(18,232){{\Large $\bm{u}$}}
\put(275,90){$t=0.6$}
\put(265,58){$t=0.5$}
\put(215,50){$t=0.4$}
\epsfbox{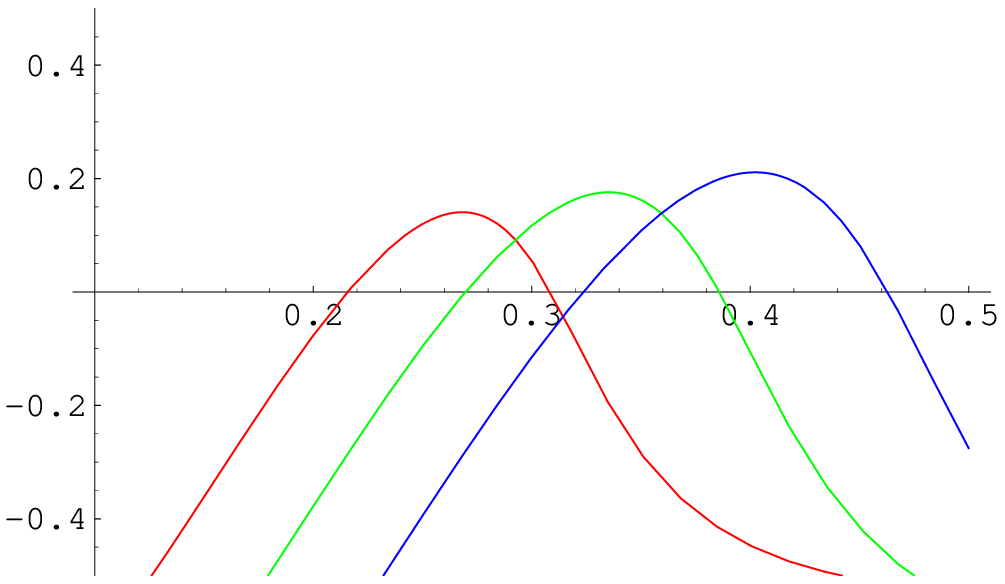}
\vspace{-1.5cm}
\caption{
Curves of $\det H=0$ in the $(v,u)$ plane for $t=0.4,0.5,0.6$.
}
\label{fig:resolhu}
\end{center}
\end{figure}

Fig.\ \ref{fig:resolhu} shows the curves of $\det H=0$ in the $(v,u)$
plane for three values of $t$.
{}From fig.\ \ref{fig:resolhu}, we see that, for $t\sim 0.5$ and $u\sim
0.2$ corresponding to the \tv, the $v$ coordinate on the
$\det H=0$ surface is $v\sim 0.35$, which is more than 50\% larger
than the value of the \tv.
In spite of this discrepancy, it seems miraculous that the
surface $\det H=0$ is rather close to the \tv.
Therefore, expecting that the discrepancy would be resolved in a more
precise analysis, let us proceed to the study of another condition,
$\bm{f}_0\propto \bm{e}_0$.

Ideally, we should test whether the two vectors $\bm{e}_0$ and
$\bm{f}_0$ become parallel at some point on the curve of the
intersections of the two surfaces $\det G=0$ and $\det H=0$.
However, though the two surfaces are separately close to the \tv,
their intersection curve is not so close to it in the present
approximation.\footnote{
For $u$ in the range $0<u<0.3$, the corresponding $t$ and $v$
coordinates on the intersection curve do not change rapidly and take
values $t\sim 0.9$ and $v\sim 0.5$. For large values of $u$, the
corresponding $v$ coordinate become outside the range $0<v<1$.
}
Therefore, we instead carry out an indirect analysis.
As $\bm{e}_0$, we take the zero-mode of $G_{ij}(\varphi)$ at a fixed
$\varphi$ which is on the surface $\det G=0$ and close to the \tv.
Concretely, we take $\varphi=(t_r,u_r,v_r)$ of
(\ref{eq:varphi_r}), and the corresponding zero-mode is
$\bm{e}_0^r=(-0.803, 0.308, 0.510)$.
Then, we plot, on the surface $\det H=0$,
the absolute value of the inner products
$|\bm{f}_1\cdot\bm{e}_0^r|$ and $|\bm{f}_2\cdot\bm{e}_0^r|$, where
$\bm{f}_1$ and $\bm{f}_2$ are the normalized eigenvectors of $H_{ij}$
corresponding to non-zero eigenvalues.
On the intersection curve, the condition $\bm{f}_0\propto \bm{e}_0$
is equivalent to $\bm{f}_1\cdot\bm{e}_0=\bm{f}_2\cdot\bm{e}_0=0$.
We are going to examine the behavior of the approximate quantities to
the two inner products on the surface $\det H=0$.

\begin{figure}[htb]
\begin{center}
\leavevmode
\epsfxsize=100mm
\put(288,65){{\Large $\bm{v}$}}
\epsfbox{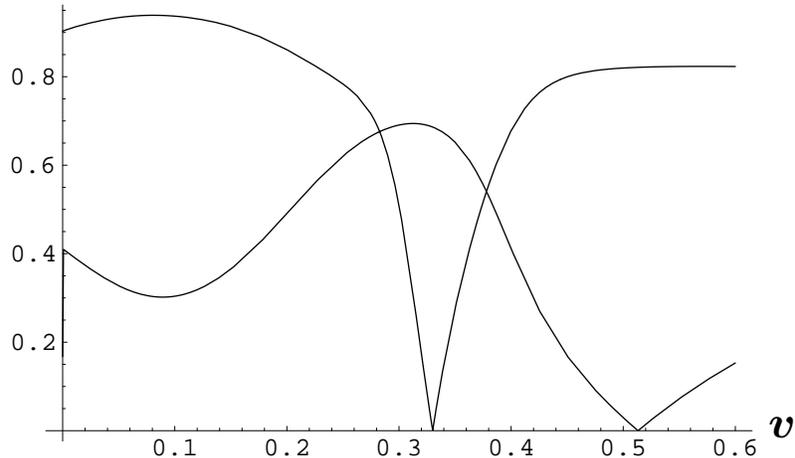}
\vspace{-2cm}
\caption{
$|\bm{f}_1\cdot\bm{e}_0^r|$ and $|\bm{f}_2\cdot\bm{e}_0^r|$ as a
function of $v$ on the surface $\det H=0$ and with $t=t_r$.
}
\label{fig:fe}
\end{center}
\end{figure}

Fig.\ \ref{fig:fe} shows these two inner products as a function of $v$
with $t$ fixed to $t_r$ and $u$ determined by $\det H=0$.
Note that one of the inner products vanishes at $v\sim 0.3$ and the
other at $v\sim 0.47$.
Moreover, the positions of these two zeros, in particular,
$v\sim 0.3$, are fairly insensitive to the choice of $\bm{e}_0$ used
for taking the inner products.
Namely, the positions of the zeros change little even if we
replace $\bm{e}_0^r$ by almost any vector.
This is ascribed to the fact that the eigenvectors $\bm{f}_1$ and
$\bm{f}_2$ change their directions rapidly by large angle in the
neighborhood of $v\sim 0.3$ and that of $v\sim 0.47$,
respectively (recall that $\bm{f}\cdot\bm{e}_0$ vanishes when
$\bm{f}$ crosses the surface perpendicular to $\bm{e}_0$).
Although the inner product $\bm{f}_{1,2}\cdot\bm{e}_0$ is not
invariant under the change of the field space coordinates $\varphi^i$,
the positions of the zeros are insensitive to the coordinate change
owing to the same property of $\bm{f}_{1,2}$.
For example, if we adopt the coordinates of \cite{MT},
$(\psi^1,\psi^2,\psi^3)=(2\,t, v/\sqrt{13},-2\,u)$, the global
behavior of the inner products differs from fig.\ \ref{fig:fe}, but
there are zeros near the corresponding points.

Therefore, the presence of the two zeros of the inner products
$\bm{f}_{1,2}\cdot\bm{e}_0^r$, in particular, the zero at $v\sim 0.3$
which is closer to the \tv, is very encouraging. It is
expected that, as we improve the approximation, the two zeros approach
to each other and to $v\sim 0.2$.

\subsection{Higher derivative terms in the BRST transformation}

We have defined the physical scalar field $T$ (\ref{eq:T}) by the
requirement that it be invariant under the BRST transformation
$\delBab$ (\ref{eq:expanddelBab}) truncated at the lowest order in the
derivatives on $C$ and $\p_\mu C_\mu$.
However, as given in (\ref{eq:expanddelB}), the full BRST
transformation contains higher derivative terms, and we have to
check whether the $\p^2 C$ and $\p^2\p_\mu C_\mu$ terms are also
cancelled in $\delB T$.
Namely, we shall test whether $Y(\varphi)$ and $Z(\varphi)$,
\begin{align}
&Y(\varphi)=\y^t + \beta_u\,\y^u + \beta_v\,\y^v ,
\\
&Z(\varphi)=\z^t + \beta_u\,\z^u + \beta_v\,\z^v ,
\end{align}
with $(\beta_u,\beta_v)$ satisfying (\ref{eq:eqforbeta}) vanish near
the \tv.

\begin{figure}[htb]
\begin{center}
\leavevmode
\epsfxsize=100mm
\put(288,155){{\Large $\bm{u}$}}
\put(22,230){{\Large $\bm{v}$}}
\epsfbox{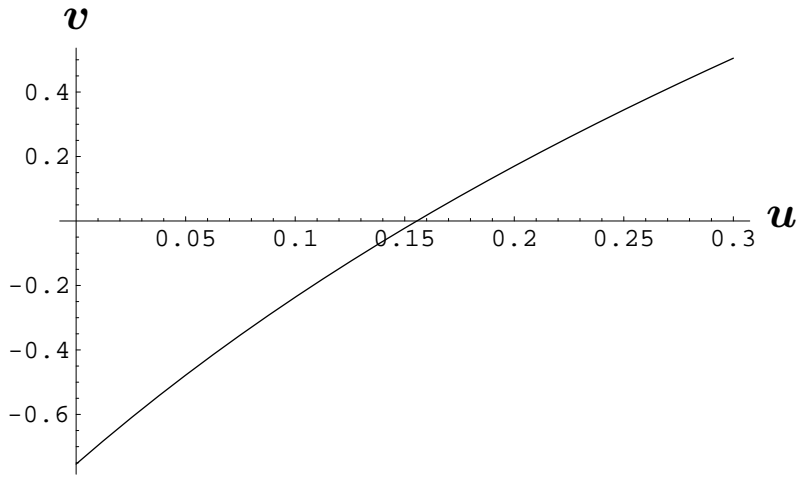}
\vspace{-2cm}
\caption{
The curve in the $(u,v)$ plane satisfying both $Y(\varphi)=0$ and
$\det G(\varphi)=0$.
}
\label{fig:zerotesty}
\end{center}
\end{figure}
\begin{figure}[htb]
\begin{center}
\leavevmode
\epsfxsize=100mm
\put(288,130){{\Large $\bm{t}$}}
\put(10,230){{\Large\bf $\bm{Z}$, $\bm{Y}$}}
\put(210,217){($0.2,0.2$)}
\put(280,210){($0.3,0.3$)}
\put(240,180){($0.4,0.4$)}
\epsfbox{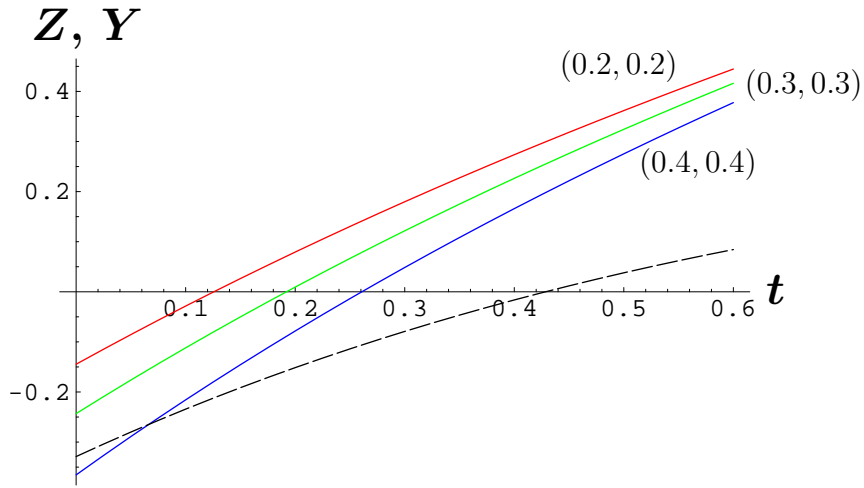}
\vspace{-2cm}
\caption{
The solid curves represent $Z(t,u,v)$ as a function of $t$ for
$(u,v)=(0.2,0,2)$, $(0.3,0,3)$ and $(0.4,0,4)$.
The dashed curve is $Y(t,u_r,v_r)$.
}
\label{fig:testyzall}
\end{center}
\end{figure}

Let us first consider $Y(\varphi)$  on the surface
$\det G(\varphi)=0$. Fig.\ \ref{fig:zerotesty} shows the curve in the
$(u,v)$ plane determined by $Y(\varphi)=0$ and $\det G(\varphi)=0$.
Note that this curve passes the point $(u,v)\sim(0.2,0.2)$, implying
that $Y(\varphi)=0$ holds near the \tv.

However, the corresponding curve for $Z(\varphi)$ is totally outside
the region of our interest in the $(u,v)$ plane. Therefore, we instead
studied the behavior of $Z(\varphi)$ as a function of $t$ for a number
of fixed values of $(u,v)$ (see fig.\ \ref{fig:testyzall}).
{}From fig.\ \ref{fig:testyzall} we see that $Z(\varphi)$ has a zero at
a rather small $t$ when the $(u,v)$ coordinate is close to that of the
\tv.

\subsection{Summary of the analysis}

In this section, we have analyzed the conditions for the absence of
the kinetic term of the BRST invariant scalar fluctuation.
The conditions can simply be
stated as (i) $G_{ij}$ and $H_{ij}$ have a common zero-mode at the \tv,
and (ii) the BRST invariant fluctuation component can be consistently
defined there. We divided the test of the conditions into many
steps and examined whether the following quantities vanish near the
\tv: $\det G$ (sec.\ 3.2), $\det H$ (sec.\ 3.3),
$|\bm{f}_{1,2}\cdot\bm{e}_0^r|$ (sec.\ 3.3), $Y$ and $Z$ (sec.\ 3.4).
Among them, we obtained fairly good results for $\det G$, $\det H$,
$|\bm{f}_{1}\cdot\bm{e}_0^r|$ (the one having a zero at $v\sim 0.3$
in fig.\ \ref{fig:fe}) and $Y$.
However, the behaviors of $|\bm{f}_{2}\cdot\bm{e}_0^r|$ and $Z$ were
not as we expected.
They are hoped to improve in a better treatment, in
particular, by taking into account the mixing with the vector and
tensor components at level two which we neglected in this paper.
We have carried out the same analysis also in the level (2,4)
approximation, and found that the results are not qualitatively
changed.

\section{Discussions}

Our analysis in this paper is incomplete in various
aspects, and further detailed studies are needed for confirming the
absence of kinetic terms.
We have to extend the space of fluctuations to incorporate
the vector and tensor components even in the level (2,6)
approximation.
This is indispensable for making clear the full BRST structure of the
fluctuations.
We have to consider the kinetic terms with more than
four derivatives. And we have to raise the order of the level
truncation.
What is more important is of course the analytical understanding of
the absence of the kinetic terms, though this must be a difficult task
since the exact solution $\Phi_0$ for the \tv\ would be
needed. Conversely, the missing kinetic terms for the fluctuations
could give a hint for finding the exact solution.
It is also an interesting question whether the CSFT expanded around
the \tv\ is a topological theory described by a BRST exact action
\cite{Hata:1984qa,Hata:1985jj,Witten:1988ze}.
We have to clarify how such an open SFT can support the
closed string sector.

\section*{Acknowledgments}
We would like to thank M.~Bando, I.~Kishimoto, T.~Kugo,
S.~Moriyama, T.~Noguchi and S.~Shinohara for valuable discussions
and useful comments.
We would also like to thank S.~Moriyama for careful reading
of the manuscript.
The work of H.~H.\ is supported in part by Grant-in-Aid for
Scientific Research from Ministry of Education, Science, Sports and
Culture of Japan (\#12640264).

\end{document}